\documentstyle[10pt,epsf,epsfig,dp_delphititle,oldlfont]{dp_delphi}
\makeindex
\def\DpPaperGroup{EP}
\def\DpPaperRef{2001-013}
\def\DpDate{1 February 2001}
\def\DpAuthors{DELPHI Collaboration}
\def\DpSubmit{(Accepted by Phys.Lett.B)}
\def\DpTitle{{Search for a fermiophobic Higgs at LEP 2}}
\def\DpComment{ }
\def\DpEMail{ }

\begin{document}
\makeatletter
\newcount\@tempcntc
\def\@citex[#1]#2{\if@filesw\immediate\write\@auxout{\string\citation{#2}}\fi
  \@tempcnta\z@\@tempcntb\m@ne\def\@citea{}\@cite{\@for\@citeb:=#2\do
    {\@ifundefined
       {b@\@citeb}{\@citeo\@tempcntb\m@ne\@citea\def\@citea{,}{\bf ?}\@warning
       {Citation `\@citeb' on page \thepage \space undefined}}%
    {\setbox\z@\hbox{\global\@tempcntc0\csname b@\@citeb\endcsname\relax}%
     \ifnum\@tempcntc=\z@ \@citeo\@tempcntb\m@ne
       \@citea\def\@citea{,}\hbox{\csname b@\@citeb\endcsname}%
     \else
      \advance\@tempcntb\@ne
      \ifnum\@tempcntb=\@tempcntc
      \else\advance\@tempcntb\m@ne\@citeo
      \@tempcnta\@tempcntc\@tempcntb\@tempcntc\fi\fi}}\@citeo}{#1}}
\def\@citeo{\ifnum\@tempcnta>\@tempcntb\else\@citea\def\@citea{,}%
  \ifnum\@tempcnta=\@tempcntb\the\@tempcnta\else
   {\advance\@tempcnta\@ne\ifnum\@tempcnta=\@tempcntb \else \def\@citea{--}\fi
    \advance\@tempcnta\m@ne\the\@tempcnta\@citea\the\@tempcntb}\fi\fi}
 
\makeatother
\begin{titlepage}
\pagenumbering{roman}
\CERNpreprint{\DpPaperGroup}{\DpPaperRef} 
\date{{\small\DpDate}} 
\title{\DpTitle} 
\address{\DpAuthors} 
\begin{shortabs} 
\noindent
\noindent
Higgs bosons predicted by the fermiophobic scenario within 
Two Higgs Doublets Models were searched for in
the data collected  by the DELPHI detector at 
centre-of-mass energies between 189 GeV and 202 GeV, 
corresponding to a total integrated luminosity of 380 pb$^{-1}$.
No signal was found and confidence limits were derived 
in the framework of possible extensions of the Standard Model Higgs sector.
\end{shortabs}
\vfill
\begin{center}
\DpSubmit \ \\ 
\DpComment \ \\
\DpEMail \ \\
\end{center}
\vfill
\clearpage
\headsep 10.0pt
\addtolength{\textheight}{10mm}
\addtolength{\footskip}{-5mm}
\begingroup
%
\newcommand{\DpName}[2]{\hbox{#1$^{\ref{#2}}$},\hfill}
\newcommand{\DpNameTwo}[3]{\hbox{#1$^{\ref{#2},\ref{#3}}$},\hfill}
\newcommand{\DpNameThree}[4]{\hbox{#1$^{\ref{#2},\ref{#3},\ref{#4}}$},\hfill}
\newskip\Bigfill \Bigfill = 0pt plus 1000fill
\newcommand{\DpNameLast}[2]{\hbox{#1$^{\ref{#2}}$}\hspace{\Bigfill}}
%
\footnotesize
\noindent
\DpName{P.Abreu}{LIP}
\DpName{W.Adam}{VIENNA}
\DpName{T.Adye}{RAL}
\DpName{P.Adzic}{DEMOKRITOS}
\DpName{I.Ajinenko}{SERPUKHOV}
\DpName{Z.Albrecht}{KARLSRUHE}
\DpName{T.Alderweireld}{AIM}
\DpName{G.D.Alekseev}{JINR}
\DpName{R.Alemany}{CERN}
\DpName{T.Allmendinger}{KARLSRUHE}
% ...
\normalsize
\endgroup
\titlefoot{Department of Physics and Astronomy, Iowa State
     University, Ames IA 50011-3160, USA
    \label{AMES}}
\titlefoot{Physics Department, Univ. Instelling Antwerpen,
     Universiteitsplein 1, B-2610 Antwerpen, Belgium \\
     \indent~~and IIHE, ULB-VUB,
     Pleinlaan 2, B-1050 Brussels, Belgium \\
     \indent~~and Facult\'e des Sciences,
     Univ. de l'Etat Mons, Av. Maistriau 19, B-7000 Mons, Belgium
    \label{AIM}}
\titlefoot{Physics Laboratory, University of Athens, Solonos Str.
     104, GR-10680 Athens, Greece
    \label{ATHENS}}
\titlefoot{Department of Physics, University of Bergen,
     All\'egaten 55, NO-5007 Bergen, Norway
    \label{BERGEN}}
\titlefoot{Dipartimento di Fisica, Universit\`a di Bologna and INFN,
     Via Irnerio 46, IT-40126 Bologna, Italy
    \label{BOLOGNA}}
\titlefoot{Centro Brasileiro de Pesquisas F\'{\i}sicas, rua Xavier Sigaud 150,
     BR-22290 Rio de Janeiro, Brazil \\
     \indent~~and Depto. de F\'{\i}sica, Pont. Univ. Cat\'olica,
     C.P. 38071 BR-22453 Rio de Janeiro, Brazil \\
     \indent~~and Inst. de F\'{\i}sica, Univ. Estadual do Rio de Janeiro,
     rua S\~{a}o Francisco Xavier 524, Rio de Janeiro, Brazil
    \label{BRASIL}}
\titlefoot{Comenius University, Faculty of Mathematics and Physics,
     Mlynska Dolina, SK-84215 Bratislava, Slovakia
    \label{BRATISLAVA}}
\titlefoot{Coll\`ege de France, Lab. de Physique Corpusculaire, IN2P3-CNRS,
     FR-75231 Paris Cedex 05, France
    \label{CDF}}
\titlefoot{CERN, CH-1211 Geneva 23, Switzerland
    \label{CERN}}
\titlefoot{Institut de Recherches Subatomiques, IN2P3 - CNRS/ULP - BP20,
     FR-67037 Strasbourg Cedex, France
    \label{CRN}}
\titlefoot{Now at DESY-Zeuthen, Platanenallee 6, D-15735 Zeuthen, Germany
    \label{DESY}}
\titlefoot{Institute of Nuclear Physics, N.C.S.R. Demokritos,
     P.O. Box 60228, GR-15310 Athens, Greece
    \label{DEMOKRITOS}}
\titlefoot{FZU, Inst. of Phys. of the C.A.S. High Energy Physics Division,
     Na Slovance 2, CZ-180 40, Praha 8, Czech Republic
    \label{FZU}}
\titlefoot{Currently at DPNC,
     University of Geneva,
     Quai Ernest-Ansermet 24, CH-1211, Geneva, Switzerland
    \label{GENEVA}}
\titlefoot{Dipartimento di Fisica, Universit\`a di Genova and INFN,
     Via Dodecaneso 33, IT-16146 Genova, Italy
    \label{GENOVA}}
\titlefoot{Institut des Sciences Nucl\'eaires, IN2P3-CNRS, Universit\'e
     de Grenoble 1, FR-38026 Grenoble Cedex, France
    \label{GRENOBLE}}
\titlefoot{Helsinki Institute of Physics, HIP,
     P.O. Box 9, FI-00014 Helsinki, Finland
    \label{HELSINKI}}
\titlefoot{Joint Institute for Nuclear Research, Dubna, Head Post
     Office, P.O. Box 79, RU-101 000 Moscow, Russian Federation
    \label{JINR}}
\titlefoot{Institut f\"ur Experimentelle Kernphysik,
     Universit\"at Karlsruhe, Postfach 6980, DE-76128 Karlsruhe,
     Germany
    \label{KARLSRUHE}}
\titlefoot{Institute of Nuclear Physics and University of Mining and Metalurgy,
     Ul. Kawiory 26a, PL-30055 Krakow, Poland
    \label{KRAKOW}}
\titlefoot{Universit\'e de Paris-Sud, Lab. de l'Acc\'el\'erateur
     Lin\'eaire, IN2P3-CNRS, B\^{a}t. 200, FR-91405 Orsay Cedex, France
    \label{LAL}}
\titlefoot{LIP, IST, FCUL - Av. Elias Garcia, 14-$1^{o}$,
     PT-1000 Lisboa Codex, Portugal
    \label{LIP}}
\titlefoot{Department of Physics, University of Liverpool, P.O.
     Box 147, Liverpool L69 3BX, UK
    \label{LIVERPOOL}}
\titlefoot{LPNHE, IN2P3-CNRS, Univ.~Paris VI et VII, Tour 33 (RdC),
     4 place Jussieu, FR-75252 Paris Cedex 05, France
    \label{LPNHE}}
\titlefoot{Department of Physics, University of Lund,
     S\"olvegatan 14, SE-223 63 Lund, Sweden
    \label{LUND}}
\titlefoot{Universit\'e Claude Bernard de Lyon, IPNL, IN2P3-CNRS,
     FR-69622 Villeurbanne Cedex, France
    \label{LYON}}
\titlefoot{Univ. d'Aix - Marseille II - CPP, IN2P3-CNRS,
     FR-13288 Marseille Cedex 09, France
    \label{MARSEILLE}}
\titlefoot{Dipartimento di Fisica, Universit\`a di Milano and INFN-MILANO,
     Via Celoria 16, IT-20133 Milan, Italy
    \label{MILANO}}
\titlefoot{Dipartimento di Fisica, Univ. di Milano-Bicocca and
     INFN-MILANO, Piazza delle Scienze 3, IT-20126 Milan, Italy
    \label{MILANO2}}
\titlefoot{IPNP of MFF, Charles Univ., Areal MFF,
     V Holesovickach 2, CZ-180 00, Praha 8, Czech Republic
    \label{NC}}
\titlefoot{NIKHEF, Postbus 41882, NL-1009 DB
     Amsterdam, The Netherlands
    \label{NIKHEF}}
\titlefoot{National Technical University, Physics Department,
     Zografou Campus, GR-15773 Athens, Greece
    \label{NTU-ATHENS}}
\titlefoot{Physics Department, University of Oslo, Blindern,
     NO-1000 Oslo 3, Norway
    \label{OSLO}}
\titlefoot{Dpto. Fisica, Univ. Oviedo, Avda. Calvo Sotelo
     s/n, ES-33007 Oviedo, Spain
    \label{OVIEDO}}
\titlefoot{Department of Physics, University of Oxford,
     Keble Road, Oxford OX1 3RH, UK
    \label{OXFORD}}
\titlefoot{Dipartimento di Fisica, Universit\`a di Padova and
     INFN, Via Marzolo 8, IT-35131 Padua, Italy
    \label{PADOVA}}
\titlefoot{Rutherford Appleton Laboratory, Chilton, Didcot
     OX11 OQX, UK
    \label{RAL}}
\titlefoot{Dipartimento di Fisica, Universit\`a di Roma II and
     INFN, Tor Vergata, IT-00173 Rome, Italy
    \label{ROMA2}}
\titlefoot{Dipartimento di Fisica, Universit\`a di Roma III and
     INFN, Via della Vasca Navale 84, IT-00146 Rome, Italy
    \label{ROMA3}}
\titlefoot{DAPNIA/Service de Physique des Particules,
     CEA-Saclay, FR-91191 Gif-sur-Yvette Cedex, France
    \label{SACLAY}}
\titlefoot{Instituto de Fisica de Cantabria (CSIC-UC), Avda.
     los Castros s/n, ES-39006 Santander, Spain
    \label{SANTANDER}}
\titlefoot{Dipartimento di Fisica, Universit\`a degli Studi di Roma
     La Sapienza, Piazzale Aldo Moro 2, IT-00185 Rome, Italy
    \label{SAPIENZA}}
\titlefoot{Inst. for High Energy Physics, Serpukov
     P.O. Box 35, Protvino, (Moscow Region), Russian Federation
    \label{SERPUKHOV}}
\titlefoot{J. Stefan Institute, Jamova 39, SI-1000 Ljubljana, Slovenia
     and Laboratory for Astroparticle Physics,\\
     \indent~~Nova Gorica Polytechnic, Kostanjeviska 16a, SI-5000 Nova Gorica, Slovenia, \\
     \indent~~and Department of Physics, University of Ljubljana,
     SI-1000 Ljubljana, Slovenia
    \label{SLOVENIJA}}
\titlefoot{Fysikum, Stockholm University,
     Box 6730, SE-113 85 Stockholm, Sweden
    \label{STOCKHOLM}}
\titlefoot{Dipartimento di Fisica Sperimentale, Universit\`a di
     Torino and INFN, Via P. Giuria 1, IT-10125 Turin, Italy
    \label{TORINO}}
\titlefoot{Dipartimento di Fisica, Universit\`a di Trieste and
     INFN, Via A. Valerio 2, IT-34127 Trieste, Italy \\
     \indent~~and Istituto di Fisica, Universit\`a di Udine,
     IT-33100 Udine, Italy
    \label{TU}}
\titlefoot{Univ. Federal do Rio de Janeiro, C.P. 68528
     Cidade Univ., Ilha do Fund\~ao
     BR-21945-970 Rio de Janeiro, Brazil
    \label{UFRJ}}
\titlefoot{Department of Radiation Sciences, University of
     Uppsala, P.O. Box 535, SE-751 21 Uppsala, Sweden
    \label{UPPSALA}}
\titlefoot{IFIC, Valencia-CSIC, and D.F.A.M.N., U. de Valencia,
     Avda. Dr. Moliner 50, ES-46100 Burjassot (Valencia), Spain
    \label{VALENCIA}}
\titlefoot{Institut f\"ur Hochenergiephysik, \"Osterr. Akad.
     d. Wissensch., Nikolsdorfergasse 18, AT-1050 Vienna, Austria
    \label{VIENNA}}
\titlefoot{Inst. Nuclear Studies and University of Warsaw, Ul.
     Hoza 69, PL-00681 Warsaw, Poland
    \label{WARSZAWA}}
\titlefoot{Fachbereich Physik, University of Wuppertal, Postfach
     100 127, DE-42097 Wuppertal, Germany
    \label{WUPPERTAL}}
\addtolength{\textheight}{-10mm}
\addtolength{\footskip}{5mm}
\clearpage
\headsep 30.0pt
\end{titlepage}
%
\pagenumbering{arabic} 
\setcounter{footnote}{0} %
\large
\def\asmz{$\alpha_s(M_Z)$}
\def\ass{\alpha_s(\sqrt{s})}
\newcommand{\kos}{\ifmmode {{\mathrm K}^{0}_{S}} \else
${\mathrm K}^{0}_{S}$\fi}
\newcommand{\kpm}{\ifmmode {{\mathrm K}^{\pm}} \else
${\mathrm K}^{\pm}$\fi}
\newcommand{\ko}{\ifmmode {{\mathrm K}^{0}} \else
${\mathrm K}^{0}$\fi}
\def\as{$\alpha_s$}
\def\asb{$\alpha_s\sp{b}$}
\def\asc{$\alpha_s\sp{c}$}
\def\asuds{$\alpha_s\sp{udsc}$}
\def\Lam{$\Lambda$}
\def\ZP{Z.\ Phys.\ {\bf C}}
\def\PL{Phys.\ Lett.\ {\bf B}}
\def\PR{Phys.\ Rev.\ {\bf D}}
\def\PRL{Phys.\ Rev.\ Lett.\ }
\def\NP{Nucl.\ Phys.\ {\bf B}}
\def\CPC{Comp.\ Phys.\ Comm.\ }
\def\NIM{Nucl.\ Instr.\ Meth.\ }
\def\Coll{Coll.,\ }
\def\Rmu{$R_3(\mu)/R_3(had)$\ }
\def\Re{$R_3(e)/R_3(had)$\ }
\def\Rmue{$R_3(\mu +e)/R_3(had)$\ }
\def\ee{$e\sp{+}e\sp{-}$}

\section {Introduction}

The spontaneous symmetry breaking mechanism is a fundamental component of the 
Standard Model (SM)  but no direct experimental evidence for the Higgs 
particles has been presented  so far.
Many of the proposed extensions of the Standard Model change the 
properties of the Higgs particles, either by the effect of new 
interactions at higher energy scales or directly by assuming a 
non-minimal Higgs sector. 
The introduction of a second Higgs doublet 
is a natural assumption and it can lead to
a scenario where a light Higgs particle with 
suppressed couplings to fermions arises \cite{2HDM_RUI}.

In the Two Higgs Doublets Models (2HDM), 
the lightest scalar Higgs boson ($h^0$) 
can be produced at LEP either in association with 
a CP-odd Higgs particle or in association with a 
$Z^0$ boson.
The decay branching ratios for the lightest
scalar Higgs change with respect to Standard Model ones
and its decay to a pair of photons 
becomes dominant in large regions of the parameter space, 
while in the Standard Model this branching ratio is $\leq 10^{-3}$.
Events with isolated photons in the final state constitute rather 
distinctive signatures of this fermiophobic scenario.

We present analyses of final states with isolated photons using
the data collected by DELPHI at centre-of-mass energies ranging between
189 GeV and 202 GeV, corresponding to a total integrated luminosity of 
about 380 pb$^{-1}$. 
In this paper we include also the results from 
an analysis of 6-jet events relevant to the 2HDM scenario.
The $h^0Z^0$ production with $h^0 \rightarrow \gamma\gamma$ 
has been investigated 
previously and interpreted in other frameworks:
an analysis of previous DELPHI data can be found in reference 
\cite{higgslisboa} and
results from other LEP experiments can be found in  \cite{opal_hgg}.
Results obtained at LEP 1 will be discussed in section \ref{sec:results}.

\section{2HDM: the fermiophobic scenario}

The Two Higgs Doublets Models (2HDM)
without explicit CP violation \cite{2HDM_RUI} 
are characterised by five physical Higgs bosons:
two neutral CP-even bosons ($h^0$, $H^0$), two charged bosons
($H^{\pm}$), and one neutral CP-odd boson ($A^0$).
The important parameters for describing the 2HDM are the angles
$\alpha$ and $\beta$, 
where $\alpha$ is the mixing angle in the neutral CP-even Higgs sector
and $\tan \beta$ is the ratio of the vacuum expectation values of 
the two Higgs doublets.
A seventh parameter is fixed in the symmetry breaking, and is related to the
masses of the vector bosons $Z^0$ and $W^\pm$ which are 
nowadays extremely well measured \cite{ZW}.

In the framework of 2HDM there are four different ways in which the
Higgs doublets can couple to fermions \cite{2HDM_LEP}. The most common 
choice is the structure assumed in the Minimal Supersymmetric extension to the
Standard Model (MSSM) \cite{MSSM} : one of the Higgs doublets couples both 
to up type quarks and to leptons, and the other doublet couples to down type 
quarks. 

In this paper a model is explored where only one of the Higgs doublets is 
allowed  to couple to fermions (model type I) \cite{2HDM_RUI}. 
The coupling of the lightest CP-even
boson, $h^0$, to a fermion pair is then proportional to $\cos \alpha$. 
If $\alpha = \frac{\pi}{2}$ this coupling vanishes and $h^0$ becomes a 
fermiophobic Higgs.

In general 2HDM, the main mechanisms for the production of neutral 
Higgs bosons at LEP are $e^+e^- \rightarrow h^0 Z^0$ 
and $e^+e^- \rightarrow   h^0 A^0 $. 
These  processes have complementary cross-sections,
proportional to $\sin^2 \delta$ and  to $\cos^2 \delta$ respectively, 
where $\delta=\alpha - \beta$.
The high $\delta$ region can be studied by analysing the
Higgs-strahlung process, while the small $\delta$ region is dominated
by the associated $h^0 A^0$ production. 
The combination of both processes leads to an interpretation of 
the results as a function of $m_{h^0}$ and $m_{A^0}$.
The region in the plane ($m_{h^0}$, $m_{A^0}$) that is relevant for
the present analyses corresponds to a band $ m_{low}<m_{A^0}+m_{h^0}<m_{high}$.
The upper constraint represents the sensitivity accessible with the present
LEP 2 integrated luminosity and centre-of-mass energies
and the lower constraint corresponds to the region excluded by previous 
analyses, namely at LEP 1.

The Higgs-Higgs interactions, namely the $h^0 H^+ H^-$ vertex, depend 
on the specific 2HDM potential. 
In fact, there are two different potentials, defined by seven parameters,
which assure no CP violation. They are referred to as potential A and 
potential B \cite{2HDM_RUI}.
These potentials are equivalent so far as the Higgs couplings
to gauge bosons and fermions are concerned. 
However, differences in the Higgs-Higgs interactions lead to 
different phenomenologies
and can alter the decay width of $h^0 \rightarrow \gamma\gamma$, for which 
the $H^+$ loop has a fundamental contribution. On the other hand,
 the relevant tree-level decays
of $A^0$ are completely independent of the chosen potential. 
The two potentials also give rise to different forbidden regions in the 
parameter space accessible at LEP. Namely, a small value of $\delta$ implies a
light $h^0$ for potential A and a small difference between
$m_{h^0}$ and $m_{A^0}$ for potential B (which is also the one 
assumed in the MSSM).

In this paper the results are interpreted for both potentials.
For Potential A, 
the branching ratio of the lightest scalar Higgs ($h^0$) to two photons,
BR($h^0 \rightarrow \gamma\gamma$), depends only on $m_{h^0}$ 
and mildly on the value of $\delta$, provided that 
$m_{H^\pm}$ is above the experimental limit of 78.6~GeV/$c^2$ \cite{lep_higgs} 
and the heavier neutral scalar Higgs boson 
($H^0$) has a mass of the order of 1 TeV/c$^2$. 
For potential B, the same branching ratio depends
also on $m_{A^0}$ and $m_{H^\pm}$, and there can be large cancellations 
between the several loop contributions for some values of these parameters. 
For higher values of $m_{H^\pm}$ (above 400~GeV/$c^2$)
or higher values of $\delta$ ($\sin^2{\delta} > 0.02$), 
there are again regions free of such cancellations. 

The dominant decay modes for $m_{h^0}< m_{Z^0}$ in the 
 fermiophobic limit
(Model I and $\alpha =\frac{\pi}{2} $) are   
$h^0 \rightarrow A^0 A^0$ (tree level) 
if $m_{h^0} > 2 m_{A^0}$ and $h^0 \rightarrow \gamma \gamma$ (one-loop)
otherwise.
The decays of $h^0$ to other boson pairs can be important when
$m_{h^0} > m_{Z^0}$, namely the one loop decay $ h^0 \rightarrow   Z^0 \gamma $
can have a BR as large as 20\% for very small $\delta$ values, while the 
decay to $WW^*$ is important for large $\delta$ values.

The tree level decay modes of the $A^0$ boson are:
$A^0 \rightarrow f \overline{f}$,  $A^0 \rightarrow Z^0 h^0$,
and $A^0 \rightarrow W^{\pm} H^{\pm}$ (when kinematicaly allowed). 
The main decay of $A^0$ is into a fermion-antifermion pair, 
namely a $b \overline{b}$ pair if $m_{A^0} > 10$ GeV. 
However, above the $Z^0 h^0$ threshold, the decay  
$A^0 \rightarrow Z^0 h^0$ dominates for all $\delta < 1.3$~rad.
Finally it should be noted that in the region of very low $\delta$ 
values ($\delta<10^{-3}$~rad) and $m_{A^0} < m_{Z^0} + m_{h^0}$, 
the $A^0$ total width is very small and $A^0$ can leave the detector 
before decaying  \cite{2HDM_RUI}. 
While for potential B, final states with invisible $A^0$ are 
important only for a small band of $m_{A^0} \sim m_{h^0}$, for
 potential A they can give rise to totally invisible final states.

The several topologies contributing to the analyses are 
summarised in table  \ref{tab:topo2}. 
For $m_{h^0}>2m_{A^0}$, the final
states will not involve photons but rather
6 b-jets or only invisible particles (stable $A^0$). 
In this region the analysis of \cite{2HDM_HA} was used together with the 
interpretation of LEP 1 data.

\begin{table}[hbt]
\begin{center}
\begin{tabular}{|l|c|c|}
\hline
 Process & Final states & Relevant mass region \\
\hline
\hline
${\rm e^+e^- \to  h^0 A^0 }$  
&${ \gamma\gamma A^0  }$(long lived)          &${\rm m_{A^0} < m_{Z^0} + m_{h^0}}$\\ 
&\raisebox{-3pt}{$\rm \gamma \gamma b\bar{b}$}&${\rm m_{h^0} + m_{A^0} > 10~~GeV/c^2}$ \\
            
\hline

${\rm e^+e^- \to h^0 A^0 \to h^0 h^0 Z^0}$ 
&\raisebox{-3pt}{${ \gamma\gamma\gamma\gamma\nu\bar{\nu} }$} 
&${\rm m_{A^0} > m_{Z^0} + m_{h^0}}$ \\
& \raisebox{-3pt}{${\rm  \gamma\gamma\gamma\gamma q\bar{q} }$}&    \\

\hline

${\rm e^+e^- \to h^0 Z^0}$ 
&\raisebox{-3pt}{${\rm \gamma\gamma \nu\bar{\nu}}$} 
&${\rm m_{h^0} < 110~~GeV/c^2}$    \\
& \raisebox{-3pt}{ ${\rm \gamma\gamma q\bar{q}}$}   &     \\


\hline
\end{tabular}
\end{center}
\caption{Topologies of the final states considered in the framework of the 
explored fermiophobic scenario in 2HDM.}
\label{tab:topo2}
\end{table}

\section{Data samples, event selection and analysis}

The analysed data from the LEP runs of 1998 and 1999 were
taken at centre-of-mass energies of 189 GeV, 192 GeV, 196 GeV, 200 GeV and 
202 GeV, with integrated luminosities of about 
153, 26, 77, 85 and 42 pb$^{-1}$, respectively. 
A detailed description of the DELPHI detector and its performance can be found
in references \cite{delphidetector1,delphidetector2}. 
The most relevant subdetectors for the present analyses were the 
electromagnetic calorimeters: the High density Projection Chamber (HPC)
in the barrel region, the Forward ElectroMagnetic Calorimeter (FEMC) 
in the endcaps and the Small angle TIle Calorimeter (STIC) 
for the very-forward region; 
the Hadronic CALorimeter (HCAL, covering polar angles down to 11 degrees), 
and the  tracking devices, namely: the Vertex Detector (VD), 
the Inner Detector (ID),
the Time Projection Chamber (TPC) and the Outer Detector (OD) in the barrel and
the Forward Chambers A and B (FCA, FCB) in the forward region.
The Vertex Detector is crucial for the determination of secondary vertices
and the tagging of $b$-quark jets and also for the identification of photons 
which convert inside the tracking system but after the VD.

The effects of experimental resolution on background and signal events were 
studied by generating Monte Carlo events and passing them through the full 
DELPHI simulation and reconstruction chain \cite{delphidetector2}.
The PYTHIA \cite{pythia} generator was used to simulate the 
background processes:
${\rm e^+ e^- \to Z^0 (N\gamma) \to q\bar{q} (N\gamma)}$,
${\rm e^+ e^- \to W^+W^-}$, 
${\rm e^+ e^- \to  W^\pm e^\mp \nu}$,
${\rm e^+ e^- \to Z^0 Z^0/\gamma^*}$,
and ${\rm e^+ e^- \to Z^0 e^+ e^-}$.
The ${\rm e^+ e^- \to Z^0 (N\gamma) \to \nu\bar{\nu} (N\gamma)}$,
    ${\rm e^+ e^- \to Z^0 (N\gamma) \to \mu\bar{\mu} (N\gamma)}$
and  ${\rm e^+ e^- \to Z^0 (N\gamma) \to \tau\bar{\tau} (N\gamma)}$
processes were generated with the KoralZ generator \cite{koralz}.
Bhabha events were generated with the BHWIDE generator 
\cite{bhwide}, ${\rm e^+ e^- \to \gamma \gamma (\gamma)}$ 
events according to 
\cite{kleiss}, and
Compton events according to \cite{compton_g}.
The two-photon (``$\gamma\gamma$'') physics events were generated with
the TWOGAM \cite{twogam} generator.

The two main backgrounds in the analysis are 
${\rm e^+ e^- \to q\bar{q} (N\gamma)}$ and
${\rm e^+ e^- \to \nu\bar{\nu} (N\gamma)}$.
The matrix-element in KoralZ generator (used for  $\nu\bar{\nu} (N\gamma)$) 
has the complete order $\alpha$ complemented with a third order 
leading-log expansion. The Pyhia generator (used for  $q\bar{q} (N\gamma)$)
was verified to be compatible with KoralZ for events with up to two visible 
photons. The absence of a complete description of the multiple photon 
radiation in MC generators may be a problem for very high luminosity
analysis. 

The analysis of events with isolated photons
was done in several steps. First a general selection was applied 
and isolated leptons, isolated photons and jets were reconstructed. 
Events with isolated leptons were removed from the analysis.

Charged particles were considered only if they had
momentum greater than 0.1 GeV/$c$ and impact parameters below 
4 cm in the transverse plane and below
4 cm /$\sin{\theta}$ in the beam direction 
($\theta$ is the polar angle, defined in relation to the beam axis).
Energy deposits in the calorimeters unassociated to charged particle
tracks were required to be above 0.3~GeV.

Isolated particles were defined by constructing double cones 
centered around the axis of the neutral cluster (charged particle track) 
with half opening angles of 
5$^{\circ}$ and 15$^{\circ}$ (5$^{\circ}$ and 25$^{\circ}$),
and requiring that the average energy density in the outer cone was below 
10 MeV/degree ( 15 MeV/degree), to assure isolation.
In the case of neutral deposits, no charged particle with more than 
250 MeV was allowed inside the inner cone.
The energy of the isolated particle was then re-evaluated as the sum of the 
energies (charged particle track momenta) inside the inner cone. 
For well identified photons or leptons, 
the above requirements were weakened:
 the external angle was allowed to be smaller 
and one energetic particle was allowed 
in the outer cone.

Photons were further required to have 
no HPC layer with more than 90\% 
of the photon electromagnetic energy. 
Alternatively, energy deposits above 3 GeV 
in the hadronic calorimeter were considered as photon candidates 
if at least 90\% 
of the deposited energy was in the first layer of the HCAL. 

Photons converting within the tracking system 
were recovered only in the non-hadronic topologies.

\subsection{Photonic final states}

Photons converting inside the tracking system, 
but after the Vertex Detector, 
are characterized by charged particle tracks and will be referred to as 
converted photons.  
Photons reaching the electromagnetic calorimeters before converting,
yielding no reconstructed charged particles tracks, will be referred to as
unconverted photons.
According to this classification, two different algorithms were applied in the
photon reconstruction and identification.

Energy deposits were considered unconverted photons if 
the following requirements were fulfilled:
\begin{itemize}

\item 
The energy was above 3 GeV. 

\item 
The polar angle of the energy deposit was inside one of the intervals 
$[20^\circ,35^\circ]$, $[42^\circ,88^\circ]$, $[92^\circ,138^\circ]$
or $[145^\circ,160^\circ]$ in order to reduce calorimeter edge effects.

\item 
No charged particle tracks were associated to the energy deposit.

\item
There was no VD track element
pointing to the 
energy deposit direction
within 3$^{\circ}$ (10$^{\circ}$) in azimuth
in the barrel (forward) region of DELPHI
(a VD track element was defined as
at least two hits in different VD layers
aligned within an azimuthal angle interval of 0.5$^{\circ}$, assuming
the charged particle track originated from the beam spot).

\item
If the polar angle of the energy deposit
was below 30$^\circ$ (above 150$^\circ$),
it had to be out of the 6 TPC $\phi$ intermodular divisions
by 2.5$^\circ$.

\end{itemize}

Photons converting after the VD in the polar angle 
range between $25^{\circ}$ and $155^{\circ}$ were recovered.
They were reconstructed with the help of the DURHAM jet 
clustering algorithm \cite{durham}.
All particles in the event, with exception 
of isolated neutral particles were clustered 
in jets, using 
as the resolution variable $y_{cut} = 0.003$.
Low multiplicity jets with less than 6 charged particles were 
treated as converted photon candidates
if they were associated to energy deposits fulfilling the
same requirements imposed on unconverted photons.


A common preselection was defined for all the photonic final states (level 1).
It was required that 
the visible energy in the polar angle region between $20^\circ$ and 
$160^\circ$ was greater than $0.1\sqrt{s}$. The
number of charged particle tracks was required to be less than 6, 
all without VD track elements.
At least two photons 
had to have energy greater than 5 GeV and polar angles 
between $25^\circ$ and $155^\circ$.
No particles (with the exception of isolated photons) with energy above 3 GeV
were allowed in the event;
no more than one photon converting in the tracking system was allowed.

Specific criteria were then 
applied to the photonic preselected sample
according to the final state topology
under study.

\subsubsection{Events with two photons and missing energy}

\begin{figure}[hbtp]
\begin{center}
\vspace{2.cm}
\vspace{-2.5cm}

\mbox{\epsfig{file=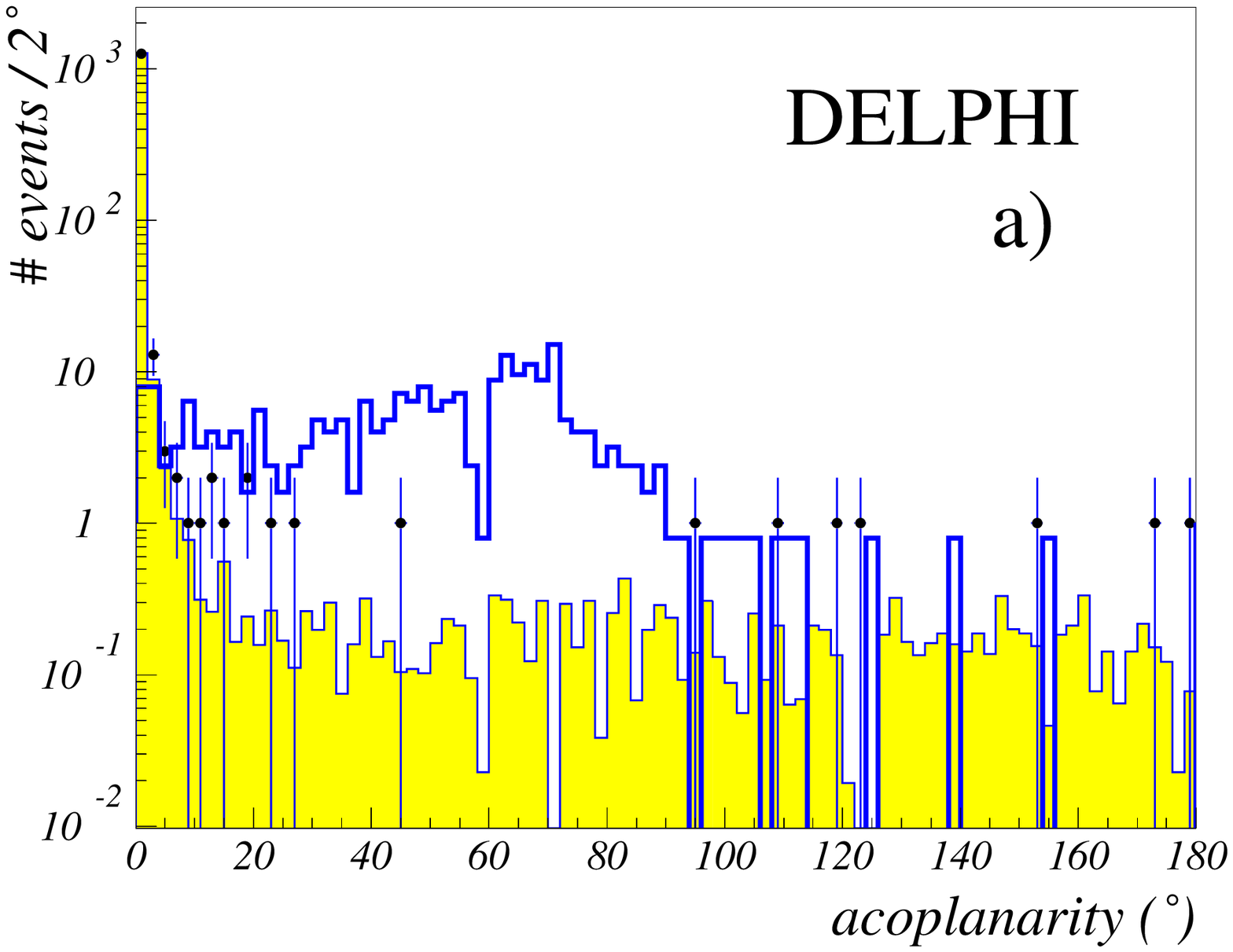,height=0.4\textheight}}

\vspace{2.5cm}

\vspace{-2.5cm}

\mbox{\epsfig{file=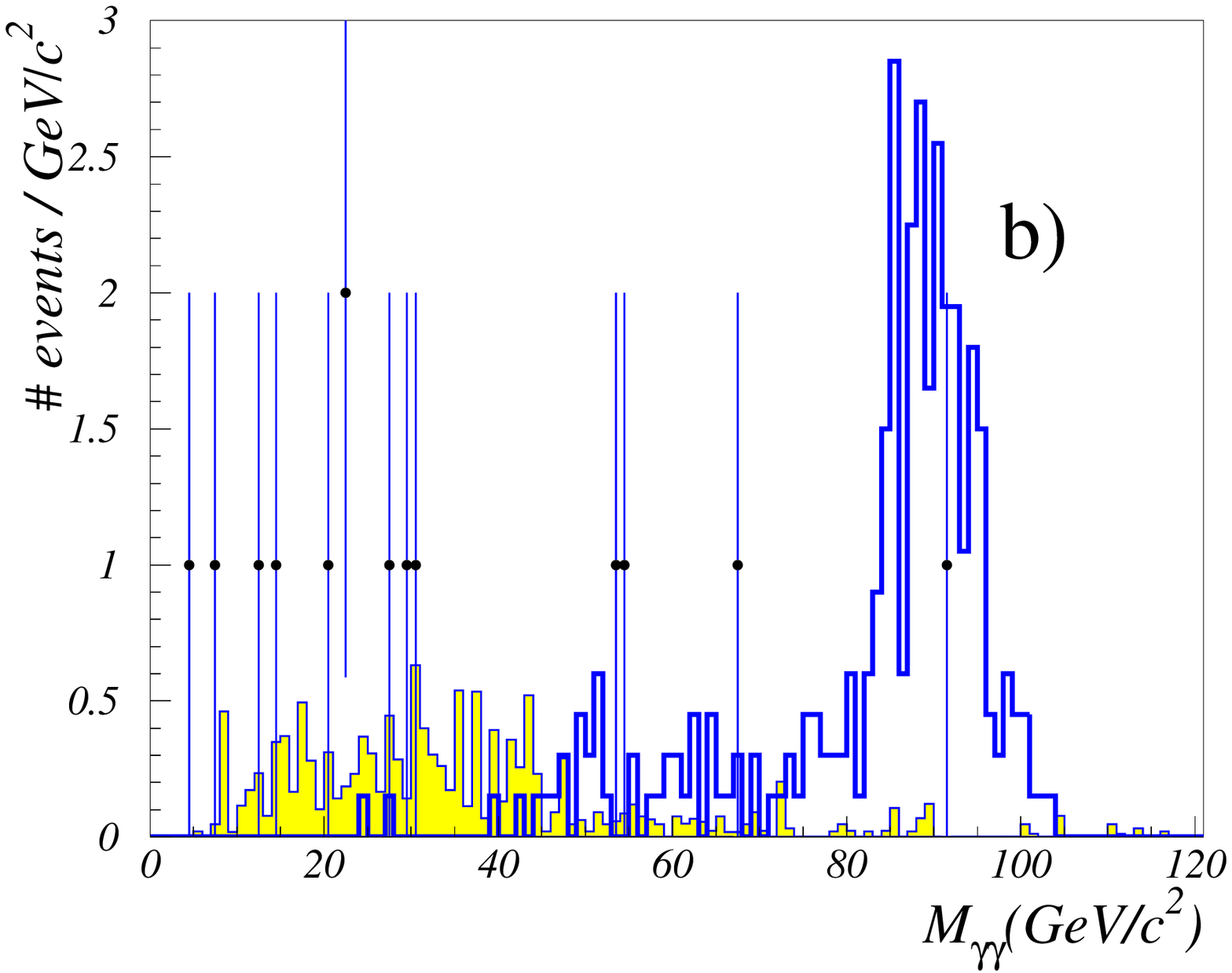,height=0.4\textheight}}

\vspace{1.cm}

\caption{Acoplanarity between the two photons at selection level 2 (a)
and invariant $\gamma\gamma$ mass at selection level 3 (b) of the two 
photon analysis.
The dots represent the data. 
The shaded area represents the expected standard model background, 
which comes mainly from the process
$e^+e^- \rightarrow Z^0 \to \nu \bar{\nu}$ with two visible ISR 
photons and from the QED process $e^+e^- \to \gamma \gamma (\gamma)$.
The darker line corresponds to signal distributions 
for a Higgs mass of 90 GeV/$c^2$, with arbitrary normalization.
The invariant mass distribution corresponds to level 3 of 
the $h^0Z^0$ selection, with
14 data events and 13$\pm$1 expected from background.}
\label{fig:mgg}
\end{center}

\end{figure}

The level 2 selection of the $\gamma \gamma + E_{miss} $ 
sample consisted of requiring events with two and only two photons. 
The acoplanarity\footnote{
acoplanarity is defined as the complement of the angle between the 
projections of the two photons in the plane perpendicular to the beam}
between the two photons in these events is 
compared to the Standard Model prediction
in figure \ref{fig:mgg}a).

Final selection criteria (level 3), aiming at 
the enhancement  
of a possible signal contribution were then imposed and consisted 
of the following conditions:

\begin{itemize}

\item 
Whenever the missing momentum was greater than
$0.1\sqrt{s}$ the polar angle of the direction
of the missing momentum was required to be greater than 
10$^{\circ}$ and less than 170$^{\circ}$ and 
no signal in the set of lead/scintillator 
counters placed between the barrel and forward 
electromagnetic calorimeters was allowed.

\item
The acoplanarity between the two photons was required to be 
greater then 10$^{\circ}$.

\item 
The sum of the energies of the two photons had to be 
lower than $0.7\sqrt{s}$.

\end{itemize}

In the case of the search for the Higgs-strahlung production,
$h^0Z^0$, with $h^0 \to \gamma \gamma$ and $Z^0 \to \nu \bar{\nu}$,
it was further required that the mass recoiling 
against the two photons was above 20 GeV/$c^2$.
The invariant masses of the photon pairs are displayed 
for these events in figure \ref{fig:mgg}b).
The background comes mainly from double radiative returns to the  
$Z^0$ with $Z\rightarrow\nu\bar{\nu}$.

The efficiencies are about $60 \%$ for both $h^0Z^0$ and
$h^0A^0$, for all centre-of-mass energies and mass ranges considered.
For $m_{h^0}$ = 90 GeV/$c^2$ and $\delta=\pi/4$, 
the number of expected signal events from $h^0Z^0$ production is 1.7.

\subsubsection{Events with four photons and missing energy}

Different criteria were imposed on the level 1 photonic sample
in order to get a wide sample of candidates 
for the associated production of $h^0 A^0$, in which the 
CP-odd boson decays to $h^0 Z^0$, the $Z^0$ going to two neutrinos.
The specific criteria for selecting
$\gamma \gamma \gamma \gamma + E_{miss}$ events (level 2), 
consisted of demanding that the events had at least three photons,
all but one between $25^\circ$ and $155^\circ$ in polar angle.
Moreover, whenever the missing momentum was greater than
$0.1\sqrt{s}$ the polar angle of the direction
of the missing momentum was required to be between 
10$^{\circ}$ and 170$^{\circ}$.

A final set of requirements was imposed in order to 
enhance a possible signal (selection level 3):

\begin{itemize}

\item 
The acoplanarity between the two most energetic photons 
had to be greater then 10$^{\circ}$.

\item 
If the missing energy was below 70 GeV, the 
missing transverse momentum had to be greater that 50 GeV/$c$.

\item
The energy of the most energetic photon had to be less than
$\sqrt{s}/2-20$ GeV.

\end{itemize}

The average efficiency of this selection is around 50$\%$.
For $m_{h^0}$ = 10 GeV/$c^2$ and $m_{A^0}$ = 120 GeV/$c^2$ and 
a $\delta=\pi/4$, 
the signal expectation is of 3.6 events,
for a total background expectation of 2.9$\pm$0.5 events, 
coming both from $Z^0\gamma\gamma$ and $\gamma\gamma$ producton.

\subsection{Final states with jets and photons}

Selection criteria were implemented to identify events with two 
jets and at least two isolated photons (level 1). Isolated photons were
reconstructed as explained in the begining of section 3. 
Their energy was further required to be above 5 GeV to avoid large 
contamination from photons coming from the hadronization.

Events were selected in the hadronic topologies if
 at least six charged particles were 
present, the visible energy in the polar angle region between 
$20^\circ$ and $160^\circ$ was greater than $0.2\sqrt{s}$
and there was 
at least one charged particle or one electromagnetic cluster
with an energy greater than 5~GeV.
All selected  charged particles and neutrals not associated to
isolated photons were forced to be clustered into two jets using the 
DURHAM jet algorithm \cite{durham}. 

For 
$ q \overline{q} \gamma \gamma$,
($ q \overline{q} \gamma \gamma \gamma \gamma$) final states 
two (at least three) photons with polar angle above $ 40^{\circ}$ and below 
$ 140^{\circ}$ were required. 
In order to improve momentum and energy resolution 
for the $q\bar{q}\gamma\gamma$ 
final states, a  kinematic 
fit \cite{pufit} imposing total energy and momentum conservation 
(with the two jets and two photons)
was performed on the selected events.
Only events with a $\chi^2$ per degree of freedom lower than 5 were
accepted. This defined the selection level 2. 
The jet-jet mass resolution at this level was 3~GeV/$c^2$.

Selection level 2 was used for the search for $h^0Z^0$.
A selection level 3 was defined
for the search for $h^0A^0$ with $A^0 \rightarrow b\bar{b}$,
in which flavour tagging was performed based on the identification of the
final state quark. 
Events with a high probability of containing a b quark 
(using the variable defined in \cite{btag1})
were thus accepted, allowing for a reduction of 50\% in the background
while keeping 90\% of the signal.

\begin{figure}[hbtp]
\begin{center}
\vspace{2.cm}
\vspace{-2.5cm}

\mbox{\epsfig{file=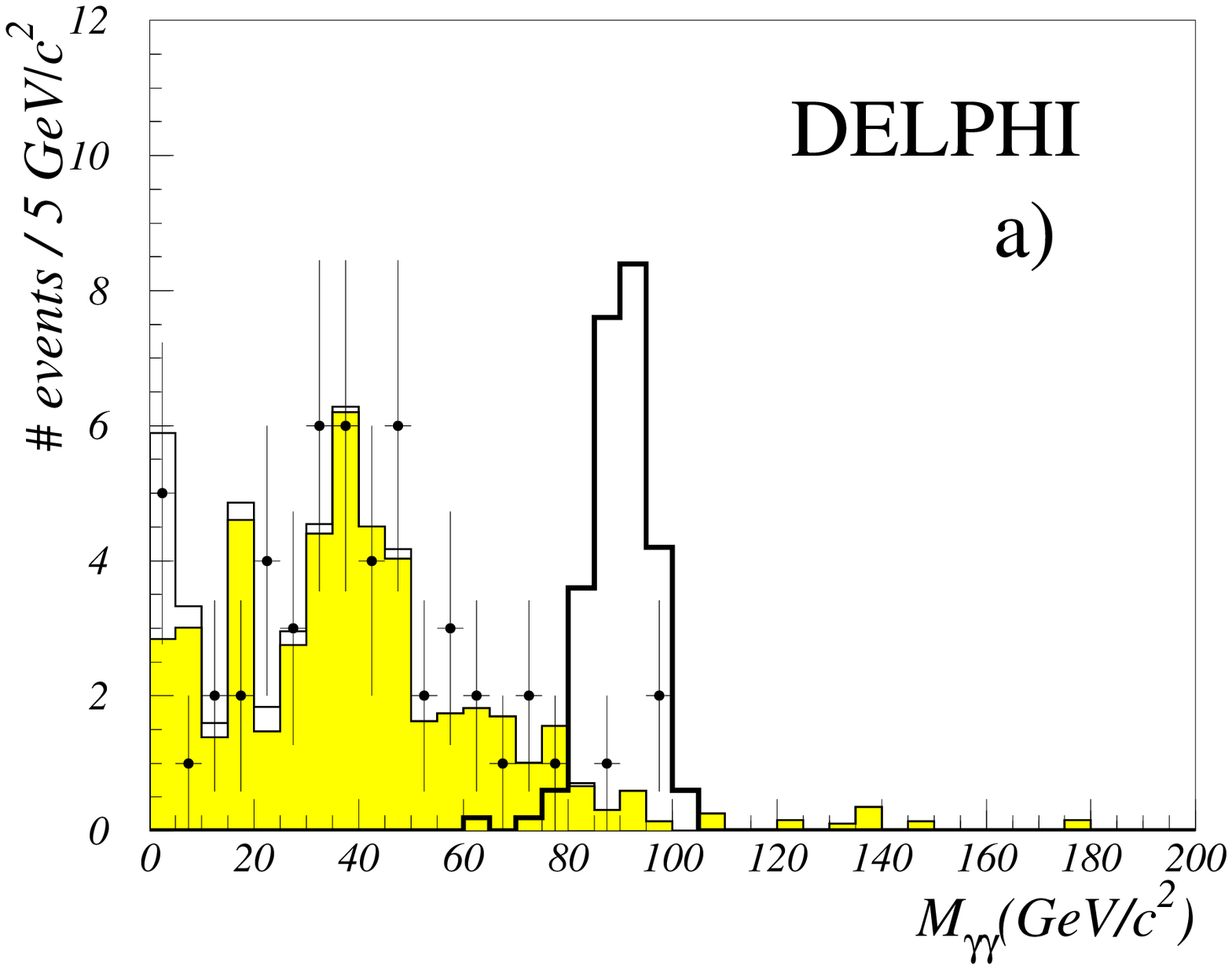,height=0.4\textheight}}

\vspace{2.5cm}

\vspace{-2.5cm}

\mbox{\epsfig{file=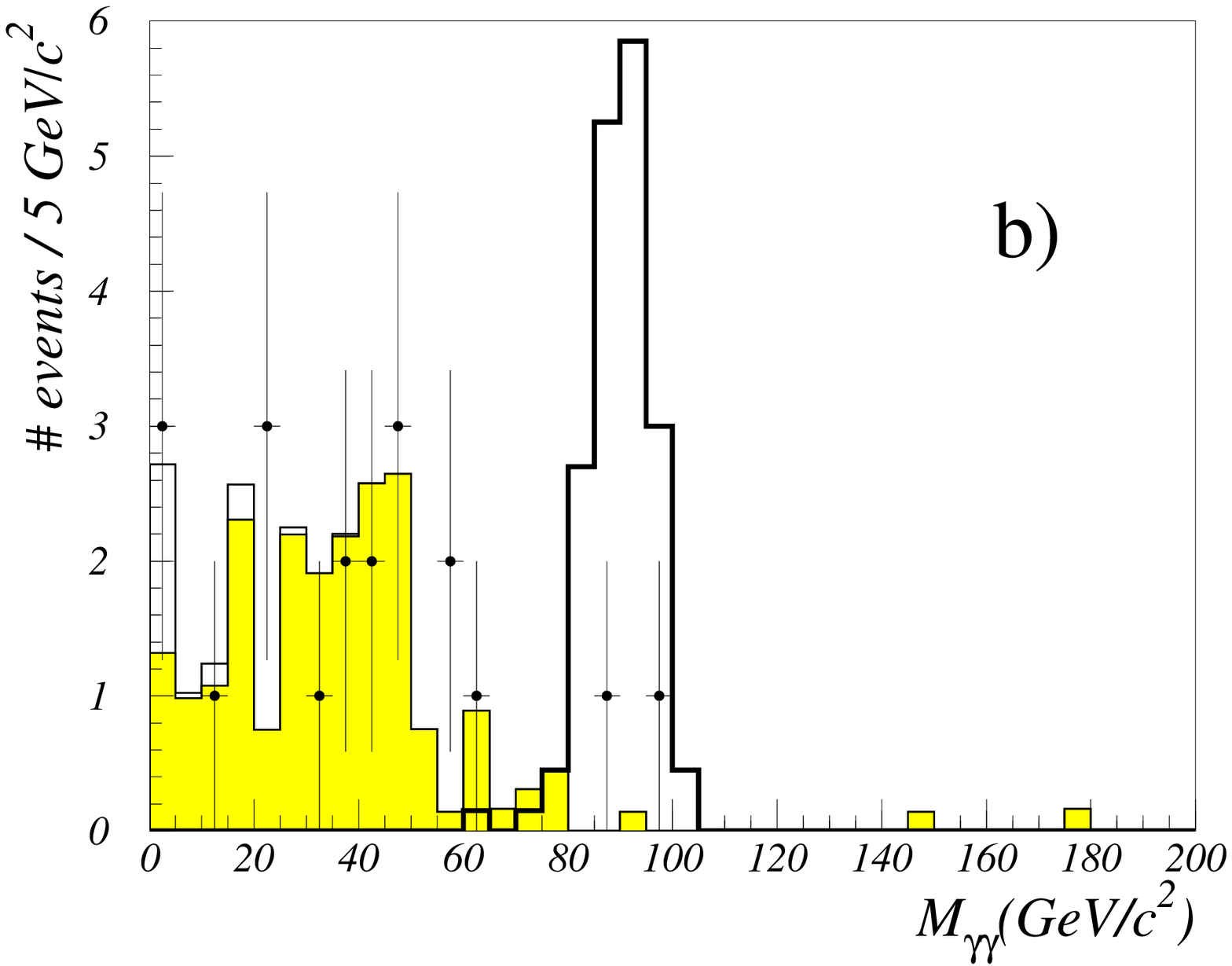,height=0.4\textheight}}

\vspace{1.cm}

\caption{Invariant $\gamma\gamma$ mass 
at selection level 2 (a) and 3 (b) of the 
$q\bar{q}\gamma\gamma$ and $b\bar{b}\gamma\gamma$ analyses.
All the data analysed are shown by dots and the shaded histograms represent
the total backgrounds. The main background contribution is 
$e^+e^- \rightarrow Z^{0*}/\gamma^* \rightarrow q\bar{q}\gamma\gamma$. The darker lines represent 
signals for a 90 GeV/$c^2$ Higgs, with arbitrary normalization.}
\label{fig:mgg_had}
\end{center}
\end{figure}

The $\gamma\gamma$ invariant masses reconstructed for events with
two jets and two photons are displayed
in figure \ref{fig:mgg_had}, both
for $h^0 Z^0$ (a) and $h^0 A^0$ (b) searches.

The average efficiencies for masses near the upper kinematic limit
are 36$\%$ and 33$\%$ for two photon events from 
$h^0Z^0$ and $h^0A^0$ production, respectively, and 30$\%$ for the 
final state with at least three photons.
These numbers correspond to expectations of 2.4 events from $h^0Z^0$
in the $q\bar{q}\gamma\gamma$ selection
and 2.1 from $h^0A^0$ in the $b\bar{b}\gamma\gamma$ selection, for
$m_{A^0}=m_{h^0}$=90 GeV/$c^2$  and $\delta=\pi/4$. 
For the final state with at least three photons and for masses of 
$m_{A^0}$=120 GeV/$c^2$ and $m_{h^0}$=10 GeV and $\delta=\pi/4$,
the signal expectation is of 5.0 events to be compared with 3.0$\pm$0.6 
background events coming mainly from $q\bar{q}\gamma\gamma\gamma$.

\section{Results\label{sec:results}}

\begin{table}[th]
\begin{center}

\begin{tabular}{|c|c|rl|rl|rl|}
\hline
{\rm Energy} & & &\multicolumn{5}{c|}{selection level} \\ 
\cline{3-8} 
{\rm GeV}&{\rm topology} &  & ~~1 &  & ~~~2 &  & ~~~3 \\
\hline
\hline
& $\gamma\gamma\gamma\gamma$   
&     &                  &91  &(85$\pm$2) & 2 & (1.5$\pm$0.4) \\
\cline{2-2} \cline{5-8}
& $\gamma\gamma$              
& 714 & (707 $\pm$ 6) &561 &(555$\pm$5) & 7 &(5.6$\pm$0.9) $(h^0 Z^0)$ \\ 
189 &     
&     &                   &    &                & 8 &(6.4$\pm$0.9) $(h^0 A^0)$ \\
\cline{2-8}
\cline{2-8}
&\raisebox{-3pt}{${\rm b\bar{b}\gamma\gamma}$} & 320 &(359$\pm$7) & 26 &(30$\pm$2) & 12 & (14$\pm$1) \\
\cline{2-8}
&${\rm q\bar{q}\gamma\gamma\gamma\gamma}$ & 25 &(27$\pm$2) & 
\multicolumn{4}{c|}{5 ~~~ (1.7$\pm$0.5)} \\
\hline
\hline
& $\gamma\gamma\gamma\gamma$  
&    &               & 12 & (13.6$\pm$0.4)& 1 & (0.15$\pm$0.05) \\
\cline{2-2} \cline{5-8}
& $\gamma\gamma$           
& 91 &(119$\pm$1)& 77 &(89$\pm$1) & 1 &(1.0$\pm$0.1) $(h^0 Z^0)$ 
\\
192 &    
&    &               &    &               & 1 &(1.1$\pm$0.1) $(h^0 A^0)$ 
\\
\cline{2-8}
\cline{2-8}
&\raisebox{-3pt}{${\rm b\bar{b}\gamma\gamma}$} & 48 &(63$\pm$2) & 2 &(2.8$\pm$0.4) & 0 & (1.0$\pm$0.2) \\
\cline{2-8}
&${\rm q\bar{q}\gamma\gamma\gamma\gamma}$ & 1 &(4.7$\pm$0.5) & 
\multicolumn{4}{c|}{0 ~~~ (0.11$\pm$0.08)} \\
\hline
\hline
& $\gamma\gamma\gamma\gamma$
&     &              &46&(40$\pm$1)       & 2  & (0.4$\pm$0.1) \\
\cline{2-2} \cline{5-8}
& $\gamma\gamma$
& 343 & (347$\pm$3) &264 &(259 $\pm$ 3) & 1 &(1.9$\pm$0.3) $(h^0 Z^0)$ \\
196 &                             
&     &              &    &                  & 1 &(2.3$\pm$0.3) $(h^0 A^0)$ 
\\
\cline{2-8}
&\raisebox{-3pt}{${\rm b\bar{b}\gamma\gamma}$} & 165 &(148$\pm$3) & 6 &(8.6$\pm$0.8) & 1 & (3.6$\pm$0.5)\\
\cline{2-8}
&${\rm q\bar{q}\gamma\gamma\gamma\gamma}$ & 8 &(10.6$\pm$0.9) & 
\multicolumn{4}{c|}{1 ~~~ (0.7$\pm$0.2)} \\
\hline
\hline
& $\gamma\gamma\gamma\gamma$   
&     &              & 43 &(42$\pm$1) & 0 & (0.6$\pm$0.2)\\
\cline{2-2} \cline{5-8}
& $\gamma\gamma$              
& 334 &(356$\pm$3)& 263 &(264$\pm$3) & 3 & (3.1$\pm$0.4) $(h^0 Z^0)$ 
\\
200 &                             
&     &              &     &              & 3 & (3.2$\pm$0.4) $(h^0 A^0)$ 
\\
\cline{2-8}
\cline{2-8}
\cline{2-8}
&\raisebox{-3pt}{${\rm b\bar{b}\gamma\gamma}$} & 149 &(153$\pm$5) & 12 &(7$\pm$1) & 6 & (2.8$\pm$0.7)\\
\cline{2-8}
&${\rm q\bar{q}\gamma\gamma\gamma\gamma}$ & 13 &(8$\pm$1) & 
\multicolumn{4}{c|}{2 ~~~ (0.4$\pm$0.2)} \\
\hline
\hline
& $\gamma\gamma\gamma\gamma$   
&     &              & 28 &(19.8$\pm$0.6)      & 0 & (0.29$\pm$0.08) \\
\cline{2-2} \cline{5-8}
& $\gamma\gamma$              
& 171 &(170$\pm$2)& 126 &(128$\pm$1)  & 2 &(1.2$\pm$0.2) $(h^0 Z^0)$ \\
202 &                             
&     &              &     &                 & 2 &(1.3$\pm$0.2) $(h^0 A^0)$ \\
\cline{2-8}
&\raisebox{-3pt}{${\rm b\bar{b}\gamma\gamma}$} & 82 &(80$\pm$3) & 7 &(3.8$\pm$0.5) & 1 & (2.1$\pm$0.4) \\
\cline{2-8}
&${\rm q\bar{q}\gamma\gamma\gamma\gamma}$ & 8 &(5.6$\pm$0.7) & 
\multicolumn{4}{c|}{0~~ (0.1$\pm$0.1)} \\
\hline
\hline
\end{tabular}
\end{center}
\caption{ Number of events passing the sets of cuts corresponding
to the selection levels described in the text for each topology 
and centre-of-mass energy. 
The MC predicted numbers of events 
and their statistical errors are displayed between parentheses. 
The second selection level of $b\bar{b}\gamma\gamma$
is the last level for the selection of $q\bar{q}\gamma\gamma$.
}
\label{tab:candidates}
\end{table}

The number of candidates at different selection levels for the relevant
topologies are given in table  \ref{tab:candidates}.
The numbers in parentheses correspond to the Standard Model
expectations which, in the case of final states with only photons, were 
corrected for trigger efficiencies (of the order of 98\% in the barrel
region of the detector and above 99\% in the forward region considered
in the analysis).
Overall, there is a reasonable agreement between data and MC expectations.

\begin{figure}[hbtp]
\begin{center}
\vspace{2.cm}
\vspace{-2.5cm}

\mbox{\epsfig{file=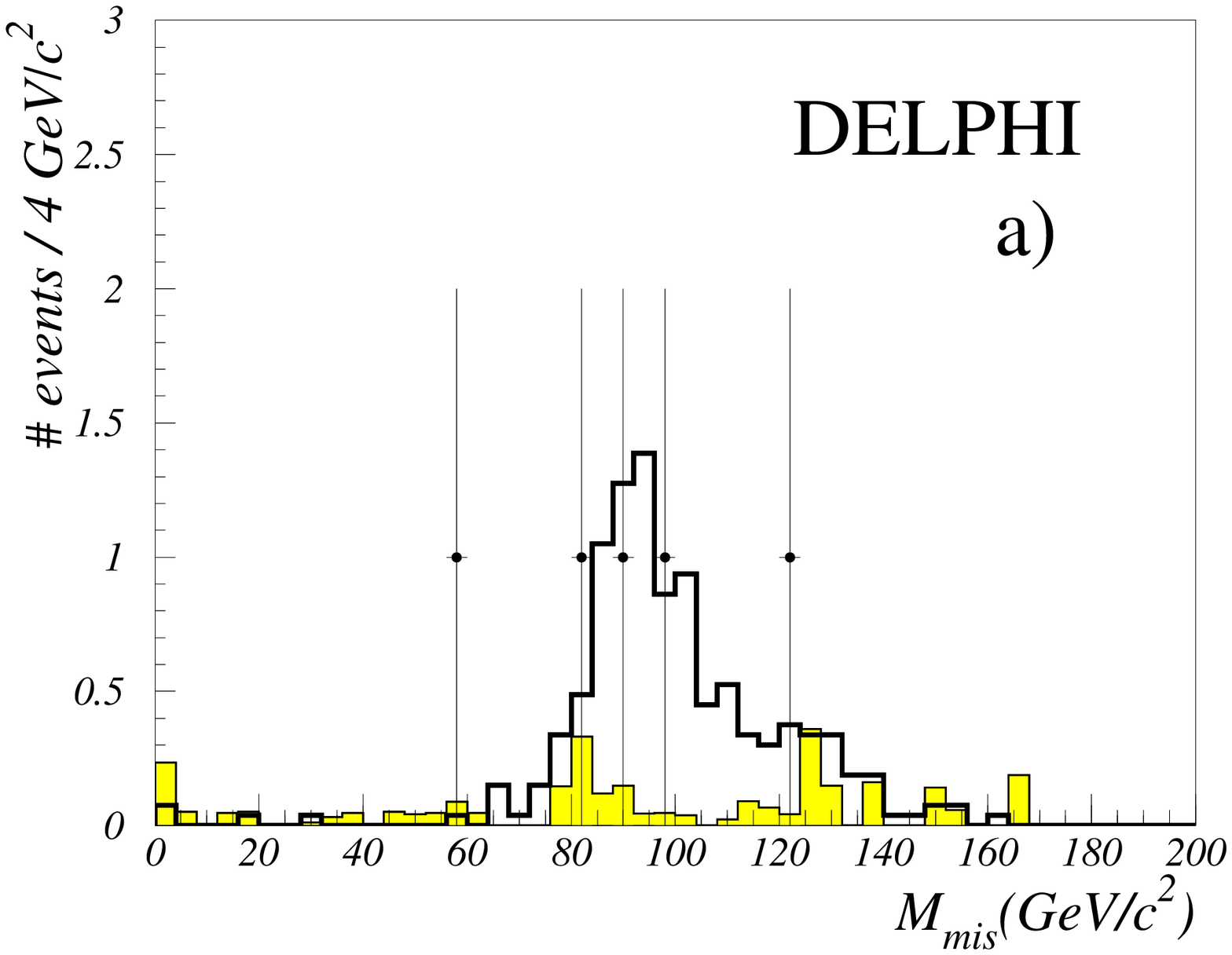,height=0.4\textheight}}

\vspace{2.5cm}
\vspace{-2.5cm}

\mbox{\epsfig{file=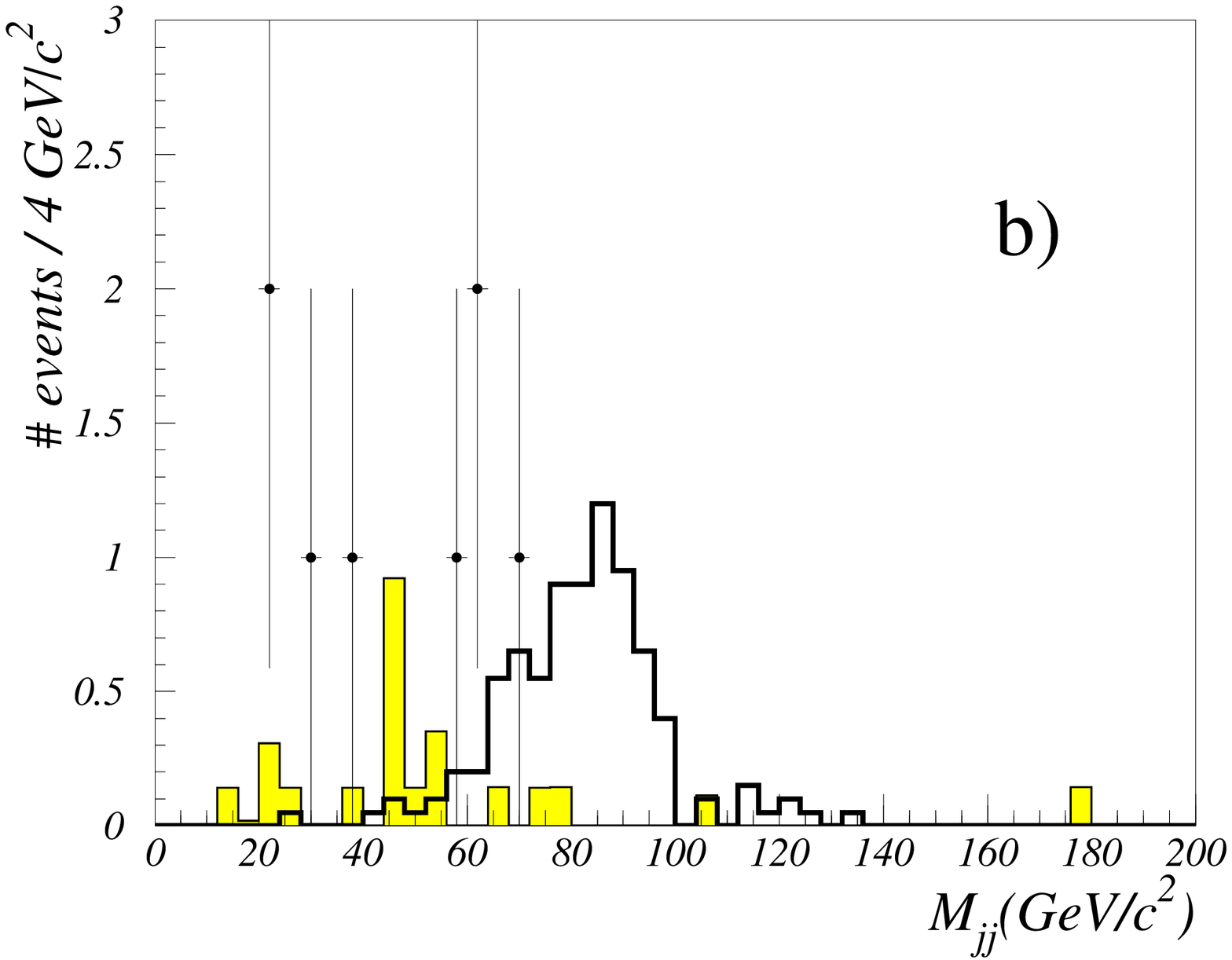,height=0.4\textheight}}

\vspace{1.cm}

\caption{Reconstructed missing mass (a) and invariant jet-jet mass (b)
for the two topologies with 4 photons, $\gamma\gamma\gamma\gamma$ 
and $q\bar{q}\gamma\gamma\gamma\gamma$ respectively.
The dots represent the data selected for all the analysed samples. 
The shaded areas represent the expected standard model background 
and the darker lines the expectations from $h^0A^0\rightarrow h^0h^0Z^0$
signals with arbitrary normalization.}
\label{fig:4fot}
\end{center}

\end{figure}

Small excesses appear in topologies with low statistics. For instance,
in the  $q\bar{q}\gamma\gamma\gamma\gamma$ final state 
at 189 GeV there is a slight excess not confirmed at higher energies.
The reconstructed $m_{Z^0}$ 
(missing mass or invariant mass of the two jets)
for events selected in the last level
of the two topologies with four photons are shown in figure \ref{fig:4fot}.
It should be remarked that the good description of the $Z^0(N\gamma)$
background has been confirmed only for final states with at
most two visible photons.

Signal selection efficiencies were calculated for each final state 
topology according
to the specific process to be studied. Several $(m_{h^0},m_{A^0})$ 
points covering the relevant parameter space were considered.
For all these masses, the width of the Higgs bosons
is smaller than the mass resolution.

These results were then combined and interpreted within the 2HDM fermiophobic 
framework 
giving limits on the cross-sections of the studied processes.
The 
Modified Frequentist Likelihood Ratio method described in \cite{Read} was used.
The method is based on the measured and expected mass distributions.
A test statistic is constructed as the ratio of the 
probability density functions of the signal 
plus background to background only hypotheses.

\subsection{Constraints from LEP 1}

The production of Higgs bosons at LEP 1 energies would have the effect of
increasing the 
$Z^0$ width. 
Since the $Z^0$ parameters are very well measured,
tight bounds can be derived on the Higgs mass.
However these results should be used with some care  \cite{moning}.
The Higgs production would change the result for the 
hadronic cross-section, which plays an important role in the fitting of the 
electroweak parameters.
In \cite{moning} a fit with more independent variables is performed,
assuming that only the 
$e^+e^- \rightarrow Z^0 \rightarrow e^+e^-$ and
$e^+e^- \rightarrow Z^0 \rightarrow \mu^+\mu^-$ 
have no contribution from new physics 
and that the new physics corrections to the other processes are not 
strongly flavour dependent. The limit thus obtained is almost model
independent and completely independent of the efficiency with which 
the new modes could be selected for the 
$e^+e^- \rightarrow Z^0 \rightarrow$ hadrons or
$e^+e^- \rightarrow Z^0 \rightarrow \tau^+\tau^-$ samples.
A  95\% Confidence Level (CL) upper limit of  6.3 MeV/$c^2$ is obtained 
for the change in the total $Z^0$ width,
which yields a limit of 149.4 pb for the production cross-section of 
unknown particles.

This cross-section limit can be used to constrain both 
$h^0 Z^{0^*}$ production  
if $\sin^2{\delta}=1$ and $h^0 A^0$ production if $\sin^2{\delta}=0$.
The first case corresponds to the exclusion of 
$m_{h^0}<$ 9~GeV/$c^2$ at 95\% CL
and the second to the exclusion of 
a band of values 
corresponding approximately to $m_{h^0}+m_{A^0} <$ 70~GeV/$c^2$ at 95\% CL. 
The intersection of the two regions 
is excluded for all $\delta$ values.

On the other hand, for processes where all the decay products are invisible,
the measurement of the $Z^0$ invisible width can be used, leading to tighter 
limits on their cross-section.
LEP 1 data \cite{ZW} leads to a  
cross-section upper limit of 67 pb at 95\% CL. 
This limit allows us to exclude the  totally invisible final state
arising from $h^0 A^0 \rightarrow A^0 A^0 A^0$ and $A^0$ stable 
(using potential A in the 2HDM fermiophobic scenario), 
excluding a band of values 
corresponding approximately to $m_{h^0}+m_{A^0} <$ 80~GeV/$c^2$ at 95\% CL.

\subsection{Constraints from 6-fermion final states}

In general 2HDM, the decay $h^0 \rightarrow A^0A^0$ is the dominant one when 
kinematically allowed. This gives rise to final states with 6-jets: 
6 $b$-jets for $h^0A$ production and 4 $b$-jets $+Z^0$ for the 
Higgs-strahlung process.

To cover this kinematic region ($m_{h^0} > 2 m_{A^0}$), the results 
from \cite{2HDM_HA} were used. In this paper, there is no dedicated analysis of
6-jet events, but it is shown that the analysis of 4-jet events (relevant for
the SM and MSSM Higgs searches) has enough sensitivity to exclude this 
region almost up to the kinematic limit.

A 95\% CL exclusion in the plane ($m_{h^0}, m_{A^0}$),
valid for all possible combinations of $\alpha$ and $\beta$ 
(all values of $\delta$), is  obtained by combining the numbers of 
expected events in $h^0Z^0$ and $h^0A^0$ channels and minimizing the CL with 
respect to $\sin\delta$ and $\cos\delta$.

\section{Limits on fermiophobic Higgs boson production}

\begin{figure}[hbtp]
\begin{center}
\vspace{-3cm}
\mbox{\epsfig{file=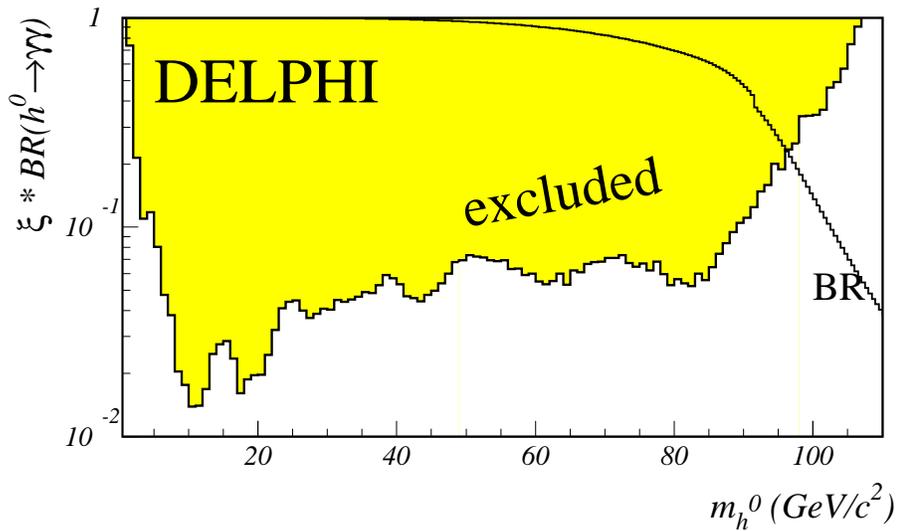,height=0.4\textheight}}
\vspace{-2mm}
\caption{95\% CL excluded region in the  
($m_{h^0}$,$\sigma(h^0Z^0)/\sigma(HZ^0)_{(SM)}$.BR$(h^0\rightarrow\gamma\gamma)$) plane,
obtained from the analysis of 
$q\bar{q}\gamma\gamma$ and $\nu\bar{\nu}\gamma\gamma$.
The BR($h^0\rightarrow\gamma\gamma$) computed in \cite{2HDM_RUI}, 
for potential A and $\sin^2\delta$=1, is also shown.}
\label{fig:xsecbr}
\end{center}
\end{figure}

In figure \ref{fig:xsecbr}, the 95\% CL limits on the 
production of a resonance $X$ in the process $Z^0X \rightarrow Z^0\gamma\gamma$
as a function of the $\gamma\gamma$ invariant mass are presented in terms 
of the product of $BR(X\rightarrow\gamma\gamma)$ and 
$\xi=\sigma(Z^0X)/\sigma(Z^0H)_{SM}$.
Here the analyses of $q\bar{q}\gamma\gamma$ and $\nu\bar{\nu}\gamma\gamma$
were used and the limit is valid for resonances with width smaller than
the  analyses mass resolution.
For a value of $\xi.BR = 1$, a limit of 107 GeV/$c^2$ 
is obtained for $m_{h^0}$. 
Also shown is 
BR($h^0 \rightarrow \gamma\gamma$) as a function of $m_{h^0}$, obtained for
potential A of the 2HDM. This branching fraction is comparable to the one 
obtained in the SM by setting to zero 
the value of the couplings of the Higgs boson to fermion pairs.

\begin{figure}[hbtp]
\begin{center}
\vspace{-3cm}
\mbox{\epsfig{file=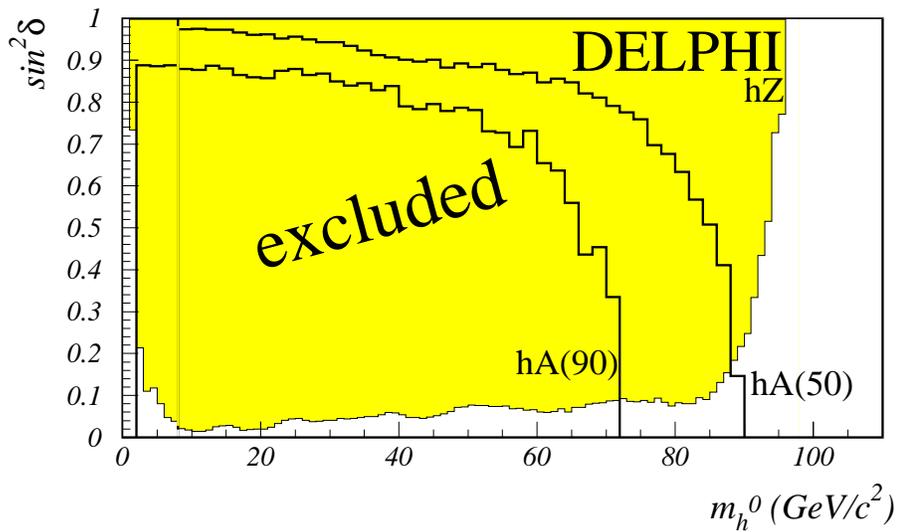,height=0.4\textheight}}
\vspace{-2mm}
\caption{95\% CL excluded region in the plane ($m_{h^0}$, $\sin^2 \delta$), 
obtained from the Higgs-strahlung final states
for potential A of 2HDM fermiophobic limit. Also shown are the 
regions excluded by the associated production process for
$m_{A^0}$= 50 GeV/$c^2$ and $m_{A^0}$ = 90 GeV/$c^2$. 
The BR($h^0 \rightarrow \gamma\gamma$) values computed in \cite{2HDM_RUI}
were used.}
\label{fig:sin2m}
\end{center}
\end{figure}

In the 2HDM scenario, $\xi$ corresponds to $\sin^2\delta$
and the BR
(which is a function of $m_{h^0}, m_{A^0}, \delta$)
 must be taken into account to determine the excluded 
($m_{h^0},\sin^2\delta$) region. 
The result for potential A
is shown in figure \ref{fig:sin2m}. The lower limit thus obtained for $m_{h^0}$
is 96 GeV/$c^2$ at 95\% CL , for $\sin^2\delta$=1. 
For small values of  $\sin^2 \delta$, the Higgs-strahlung cross-section
vanishes but an exclusion region can be obtained from the $h^0 A^0$
associated production. Such 95\% CL exclusion regions are shown for
two different $A^0$ masses. For $m_{A^0} < 60$~GeV/$c^2$, 
$m_{h^0} < 9$~GeV/$c^2$ is excluded by the $Z^0$ width measurements.

\begin{figure}[hbtp]
\begin{center}
\mbox{\epsfig{file=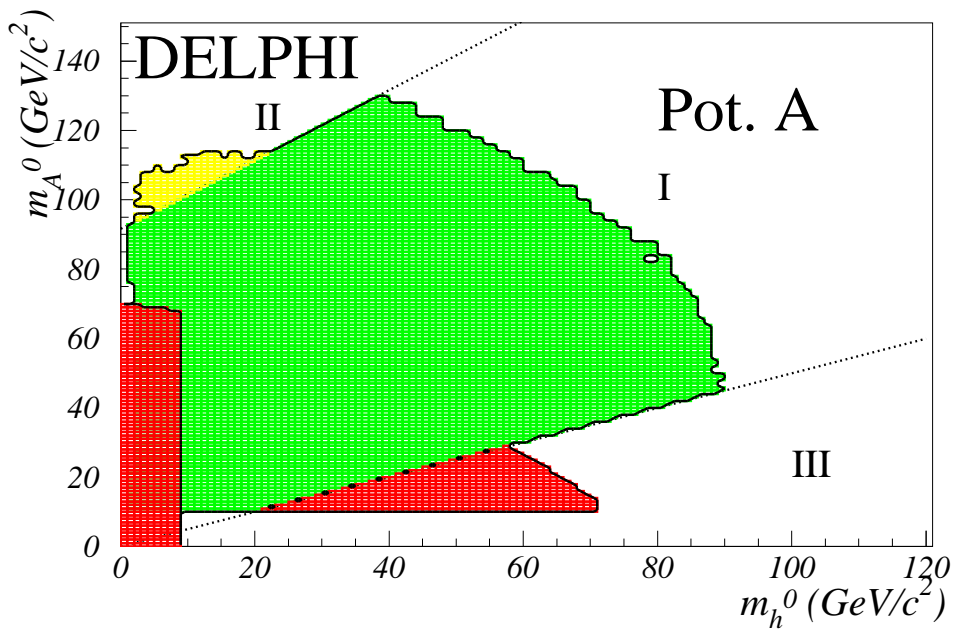,height=0.4\textheight}}
\mbox{\epsfig{file=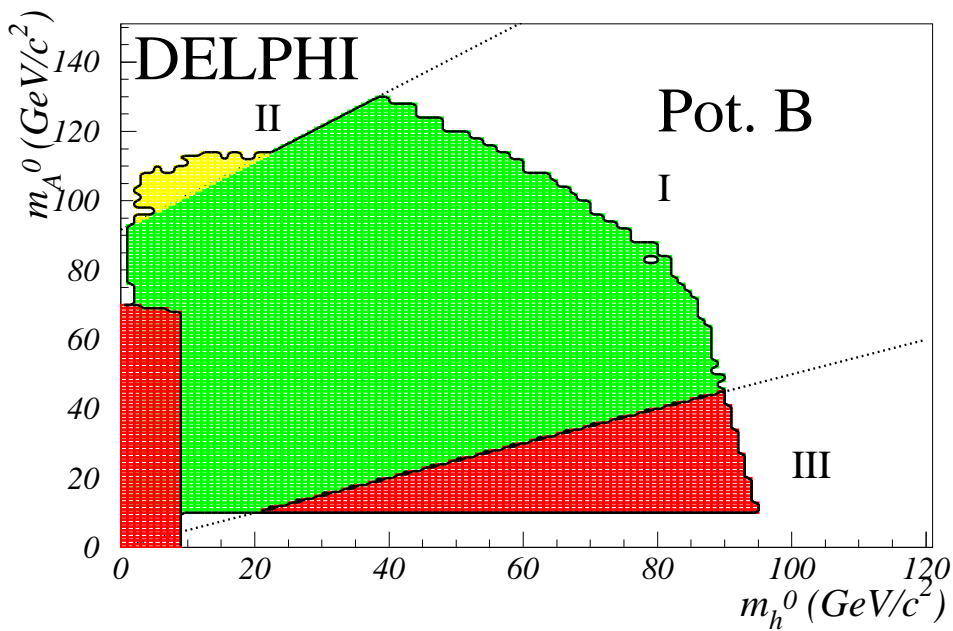,height=0.4\textheight}}

\caption{95 \% CL excluded region in the plane $(m_{h^0}$,$m_{A^0})$,
in the framework of Potential A (upper plot) and of Potential B (lower plot).
The exclusion is valid for all $\delta$ values and is obtained by combining 
the Higgs-strahlung and the associated production processes.
Region I corresponds to the decay modes $h^0 \to \gamma \gamma$ and 
$A^0\to b \bar{b}$ (or $A^0$ long-lived, for Potential A).
Region II corresponds to $A^0 \rightarrow h^0Z^0$, from the 
Higgs-strahlung and the two final states with 4 photons.
Region III corresponds to $h^0 \to A^0 A^0$ and  $A^0\to b \bar{b}$
(taken from ref. \cite{2HDM_HA}). 
For Potential A and very small $\delta$ values, 
$A^0$ is stable and the limit for all $\delta$ comes 
from the $Z^0$ invisible width measurement.
The dark band in the low $m_{h^0}$ ($<9$~~GeV/$c^2$) region represents
the limit from the total $Z^0$ width.}
\label{fig:mhmaAB}
\end{center}
\end{figure}

Due to the complementarity of the $h^0 Z^0$ and $h^0 A^0$ cross-sections, 
regions of
the plane ($m_{h^0}$,$m_{A^0}$) can be excluded at 95 \% CL
for all $\delta$ values. Figures \ref{fig:mhmaAB}(a) and \ref{fig:mhmaAB}(b)
show the excluded regions in this plane in the framework of potentials 
A and B, respectively.
In both plots, Region I corresponds to the zone where 
$h^0\rightarrow \gamma\gamma$ and $A^0 \rightarrow b\bar{b}$. 
In the case of potential A, results of $\gamma\gamma A^0$(stable) must 
also be taken into account, however, the $\gamma\gamma$ final state gives 
stronger limits and the unexcluded region is still 
defined by the $b\bar{b}\gamma\gamma$ search.
In Region II, corresponding to $h^0\rightarrow \gamma\gamma$ and 
$A^0 \rightarrow h^0Z^0$, both $\gamma\gamma\gamma\gamma$ and 
$q\bar{q}\gamma\gamma\gamma\gamma$ are considered, together with the 
Higgs-strahlung process.
In Region III, corresponding to $h^0\rightarrow A^0 A^0$ and 
$A^0 \rightarrow b\bar{b}$
giving rise to 6-jet final states, the 95\% CL limits on
($m_{h^0},m_{A^0}$) from \cite{2HDM_HA}
are used. In the case of potential A, the $A^0$ boson can be stable and 
the limits from 
the $Z^0$ invisible width provide the most conservative exclusion region. 
As discussed previously, the measurement of the $Z^0$ width at LEP 1
allows the exclusion of a band of values for low masses of both $h^0$ and $A^0$
which is common for both potentials, this is 
also indicated in figures \ref{fig:mhmaAB}(a) and \ref{fig:mhmaAB}(b).

\section{Conclusions}

DELPHI data corresponding to a total integrated luminosity of 380 pb$^{-1}$,
at centre-of-mass energies between 189 GeV and 202 GeV, 
have been analysed and 
a search for a neutral Higgs boson with predominantly non-fermionic couplings
was performed.
The final states  
$\gamma\gamma$, 
$\gamma\gamma\gamma\gamma$,              
${\rm b\bar{b}\gamma\gamma}$,
${\rm q\bar{q}\gamma\gamma}$ and 
${\rm q\bar{q}\gamma\gamma\gamma\gamma}$ 
were considered.
A large region of the parameter space in a 2HDM fermiophobic scenario
was excluded.

\subsection*{Acknowledgements}
  We would like to thank L.Br{\" u}cher and R.Santos for very interesting 
and long discussions in exploring the 2HDM fermiophobic scenario. 

 We are greatly indebted to our technical 
collaborators, to the members of the CERN-SL Division for the excellent 
performance of the LEP collider, and to the funding agencies for their
support in building and operating the DELPHI detector.\\
We acknowledge in particular the support of \\
Austrian Federal Ministry of Education, Science and Culture,
GZ 616.364/2-III/2a/98, \\
FNRS--FWO, Flanders Institute to encourage scientific and technological 
research in the industry (IWT), Belgium,  \\
FINEP, CNPq, CAPES, FUJB and FAPERJ, Brazil, \\
Czech Ministry of Industry and Trade, GA CR 202/96/0450 and GA AVCR A1010521,\\
Danish Natural Research Council, \\
Commission of the European Communities (DG XII), \\
Direction des Sciences de la Mati$\grave{\mbox{\rm e}}$re, CEA, France, \\
Bundesministerium f$\ddot{\mbox{\rm u}}$r Bildung, Wissenschaft, Forschung 
und Technologie, Germany,\\
General Secretariat for Research and Technology, Greece, \\
National Science Foundation (NWO) and Foundation for Research on Matter (FOM),
The Netherlands, \\
Norwegian Research Council,  \\
State Committee for Scientific Research, Poland, 2P03B06015, 2P03B11116 and
SPUB/P03/DZ3/99, \\
JNICT--Junta Nacional de Investiga\c{c}\~{a}o Cient\'{\i}fica 
e Tecnol$\acute{\mbox{\rm o}}$gica, Portugal, \\
Vedecka grantova agentura MS SR, Slovakia, Nr. 95/5195/134, \\
Ministry of Science and Technology of the Republic of Slovenia, \\
CICYT, Spain, AEN96--1661 and AEN96-1681,  \\
The Swedish Natural Science Research Council,      \\
Particle Physics and Astronomy Research Council, UK, \\
Department of Energy, USA, DE--FG02--94ER40817. \\

\newpage
             

\newpage

\newpage

\newpage

\newpage

\newpage

\newpage


\newpage

\end{document}